\documentclass[final,5p,times,twocolumn]{elsarticle}

\usepackage{amssymb}
\usepackage{amsmath}
\usepackage{times}
\usepackage{subfigure}
\usepackage{lineno}

\usepackage{hyperref}

\usepackage{booktabs} 

\journal{Journal}

\begin{document}

\begin{frontmatter}



\title{Toward Open Science in the AEC Community: An Ecosystem for Sustainable Digital Knowledge Sharing and Reuse}


\author[inst1]{Ruoxin Xiong}

\affiliation[inst1]{organization={Construction Management, College of Architecture \& Environmental Design},%Department and Organization
            addressline={Kent State University}, 
            city={Kent},
            postcode={44240}, 
            state={OH},
            country={USA}}

\author[inst2]{Yanyu Wang\corref{cor1}}
\cortext[cor1]{Corresponding author}
\ead{yanyuwang@lsu.edu}
\affiliation[inst2]{organization={Bert S. Turner Department of Construction Management},%Department and Organization
            addressline={Louisiana State University}, 
            city={Baton Rouge},
            postcode={70803}, 
            state={LA},
            country={USA}}

\author[inst3]{Jiannan Cai}
\affiliation[inst3]{organization={School of Civil \& Environmental Engineering, and Construction Management},%Department and Organization
            addressline={The University of Texas at San Antonio}, 
            city={San Antonio},
            postcode={78249}, 
            state={TX},
            country={USA}}

\author[inst4]{Kaijian Liu}
\affiliation[inst4]{organization={Department of Civil, Environmental, and Ocean Engineering},%Department and Organization
            addressline={Stevens Institute of Technology}, 
            city={Hoboken},
            postcode={07030}, 
            state={NJ},
            country={USA}}

\author[inst5]{Yuansheng Zhu}
\affiliation[inst5]{organization={Department of Computing and Information Sciences},%Department and Organization
            addressline={Rochester Institute of Technology}, 
            city={Rochester},
            postcode={14623}, 
            state={NY},
            country={USA}}

\author[inst6]{Pingbo Tang}
\affiliation[inst6]{organization={Department of Civil and Environmental Engineering},%Department and Organization
            addressline={Carnegie Mellon University}, 
            city={Pittsburgh},
            postcode={15213}, 
            state={PA},
            country={USA}}

\author[inst7]{Nora El-Gohary}
\affiliation[inst7]{organization={Department of Civil and Environmental Engineering},%Department and Organization
            addressline={University of Illinois at Urbana-Champaign}, 
            city={Urbana},
            postcode={61801}, 
            state={IL},
            country={USA}}

\author[inst8]{George Edward Gibson Jr.}
\affiliation[inst8]{%Department and Organization
            addressline={National Academy of Construction}, 
            city={Austin},
            postcode={78712}, 
            state={TX},
            country={USA}}

% \affiliation[inst1]{organization={Department One},%Department and Organization
%             addressline={Address One}, 
%             city={City One},
%             postcode={00000}, 
%             state={State One},
%             country={Country One}}

% \author[inst2]{Author Two}
% \author[inst1,inst2]{Author Three}

% \affiliation[inst2]{organization={Department Two},%Department and Organization
%             addressline={Address Two}, 
%             city={City Two},
%             postcode={22222}, 
%             state={State Two},
%             country={Country Two}}

\begin{abstract}
The Architecture, Engineering, and Construction (AEC) industry is undergoing rapid digital transformation, producing diverse digital assets such as datasets, computational models, use cases, and educational materials across the built environment lifecycle. However, these resources are often fragmented across repositories and inconsistently documented, limiting their discoverability, interpretability, and reuse in research, education, and practice. This study introduces \textsc{OpenConstruction}, a community-driven open-science ecosystem that aggregates, organizes, and contextualizes openly accessible AEC digital resources. The ecosystem is structured into four catalogs, including datasets, models, use cases, and educational resources, supported by consistent descriptors, curator-led validation, and transparent governance. As of December 2025, the platform hosts 94 datasets, 65 models, and a growing collection of use cases and educational materials. Two case studies demonstrate how the ecosystem supports benchmarking, curriculum development, and broader adoption of open-science practices in the AEC sector. The platform is publicly accessible at \href{https://www.openconstruction.org/}{https://www.openconstruction.org/}.
\end{abstract}


% \begin{highlights}
% \end{highlights}

\begin{keyword}
open science \sep knowledge reuse \sep digital ecosystem \sep AEC sector \sep knowledge sharing
% %% PACS codes here, in the form: \PACS code \sep code
% %% MSC codes here, in the form: \MSC code \sep code
% %% or \MSC[2008] code \sep code (2000 is the default)
\end{keyword}

\end{frontmatter}

% \linenumbers

\section{Introduction}
\label{sec:intro}
Digital technologies are transforming how information is captured, managed, and applied across the Architecture, Engineering, and Construction (AEC) lifecycle. Advances in sensing, simulation, automation, and data-driven modeling generate extensive digital resources with the potential to support systematic analysis, cumulative learning, and the development of generalizable AEC knowledge \cite{wang2023characterizing, abioye2021artificial, xiong2025openconstruction}.

However, digital resources in AEC remain widely dispersed across project-specific workflows, documented with inconsistent metadata and incomplete contextual information \cite{bosfield2025open, hazeem2024fragmented}. This fragmentation, characteristic of the sector’s decentralized and project-based practices, severely limits the ability to transfer insights between projects and organizations \cite{elkhidir2025toward}. As a result, many data-driven efforts remain isolated, constraining cumulative learning and limiting the broader impact of digital innovations in the AEC domain \cite{jaskula2024common}.

Three challenges contribute to this gap. First, digital resources originate from diverse sensing modalities, workflows, and collection conditions, producing heterogeneous and often incompatible representations of AEC environments. Second, computational models differ in assumptions, input–output structures, and evaluation protocols, and the absence of coordinated metadata limits transparency and cross-study benchmarking. Third, project-level insights and methodological practices are seldom formalized in reusable forms, limiting adaptation and accumulation of generalizable knowledge across the domain. Addressing these challenges requires a cyberinfrastructure that can systematically organize heterogeneous resources, define clear governance structures, and support their accumulation and reuse at scale \cite{alavi2001knowledge, wenger2002cultivating}.

This study introduces \textsc{OpenConstruction}, a community-driven open-science ecosystem designed to adapt the principles of Findable, Accessible, Interoperable, and Reusable (FAIR) \cite{wilkinson2016fair} to AEC digital knowledge. Focusing exclusively on openly accessible resources, the platform structures its ecosystem into four coordinated catalogs: 1) the dataset catalog characterizes the content, modality, and domain focus of available data; 2) the model catalog documents computational approaches used to address domain-specific problems within the community; 3) the use-case catalog documents real-world AEC projects that employ advanced digital solutions, providing practice-oriented examples that conceptually complement the dataset and model catalogs; and 4) the educational-resource catalog curates open textbooks and training materials that support AEC education in areas aligned with this ecosystem’s focus. Through consistent descriptors, curator-led validation, and transparent governance, the ecosystem offers a coherent and accessible view of heterogeneous digital assets, enhancing discoverability, comparability, and reuse.

The remainder of this paper is organized as follows. Section 2 reviews current digitalization efforts in AEC and motivates the need for open-science cyberinfrastructure. Section 3 introduces the conceptual framework, and Section 4 describes its implementation workflow and platform services. Section 5 presents the current composition and usage of the catalogs, Section 6 discusses implications and remaining challenges for AEC open-science cyberinfrastructure, and Section 7 concludes with future directions.

\section{Background}
This section reviews the current state of digitalization and the need for open-science cyberinfrastructure in the AEC domain.

\subsection{Digitalization and Knowledge Fragmentation in AEC}
\label{sec:background:fragmentation}
Digitalization across the AEC domain has expanded the availability of geometric, semantic, sensing, and process data supporting Machine learning (ML) and artificial intelligence (AI)-driven analyses of design intent, system performance, and site operations \cite{abioye2021artificial}. Despite this growth, digital outputs remain fragmented. Project workflows prioritize short-term coordination rather than long-term knowledge curation, limiting the preservation, validation, and cumulative use of digital resources \cite{wang2023characterizing,bosfield2025open}. Standards such as IFC and openBIM improve information exchange during project delivery but do not provide mechanisms for semantic preservation, reproducibility, or cross-project comparability \cite{schlenger2025reference}. Similarly, existing data repositories are developed independently and rely on heterogeneous metadata structures, preventing consistent interpretation and reuse \cite{xiong2025openconstruction}. Consequently, valuable digital assets often remain isolated within project-specific environments, inhibiting the accumulation of generalizable AEC knowledge.

\subsection{Open Science Principles and Cyberinfrastructure Needs}
Coordinated open-science frameworks in other scientific fields demonstrate that transparent standards for validation and reuse are essential for cumulative knowledge development \cite{alavi2001knowledge, wenger2002cultivating, benson2012genbank}. Large-scale infrastructures such as GenBank and the Materials Project operationalize these principles through standardized metadata, persistent identifiers, and curator oversight, enabling reproducible research at scale \cite{benson2012genbank,jain2013commentary}. Central to these efforts are the FAIR principles, Findable, Accessible, Interoperable, and Reusable, which provide a foundation for sustained management and reuse of digital artifacts \cite{wilkinson2016fair}. In AEC, open practices are emerging but remain limited. Common Data Environments support project-level coordination but not long-term stewardship or cross-context comparability \cite{schlenger2025reference}. Existing public repositories prioritize data availability yet lack mechanisms to integrate datasets, computational models, and applied contexts within unified metadata frameworks \cite{wang2023characterizing}. Recent studies highlight the need for domain-specific cyberinfrastructure capable of relating heterogeneous AEC digital resources and enabling verifiable and reusable knowledge generation \cite{xiong2025openconstruction,bosfield2025open}.

\section{Conceptual Framework of the \textsc{OpenConstruction}}
\label{sec:conceptual}
\begin{figure}[htbp]
    \centering
    \includegraphics[width=0.98\linewidth]{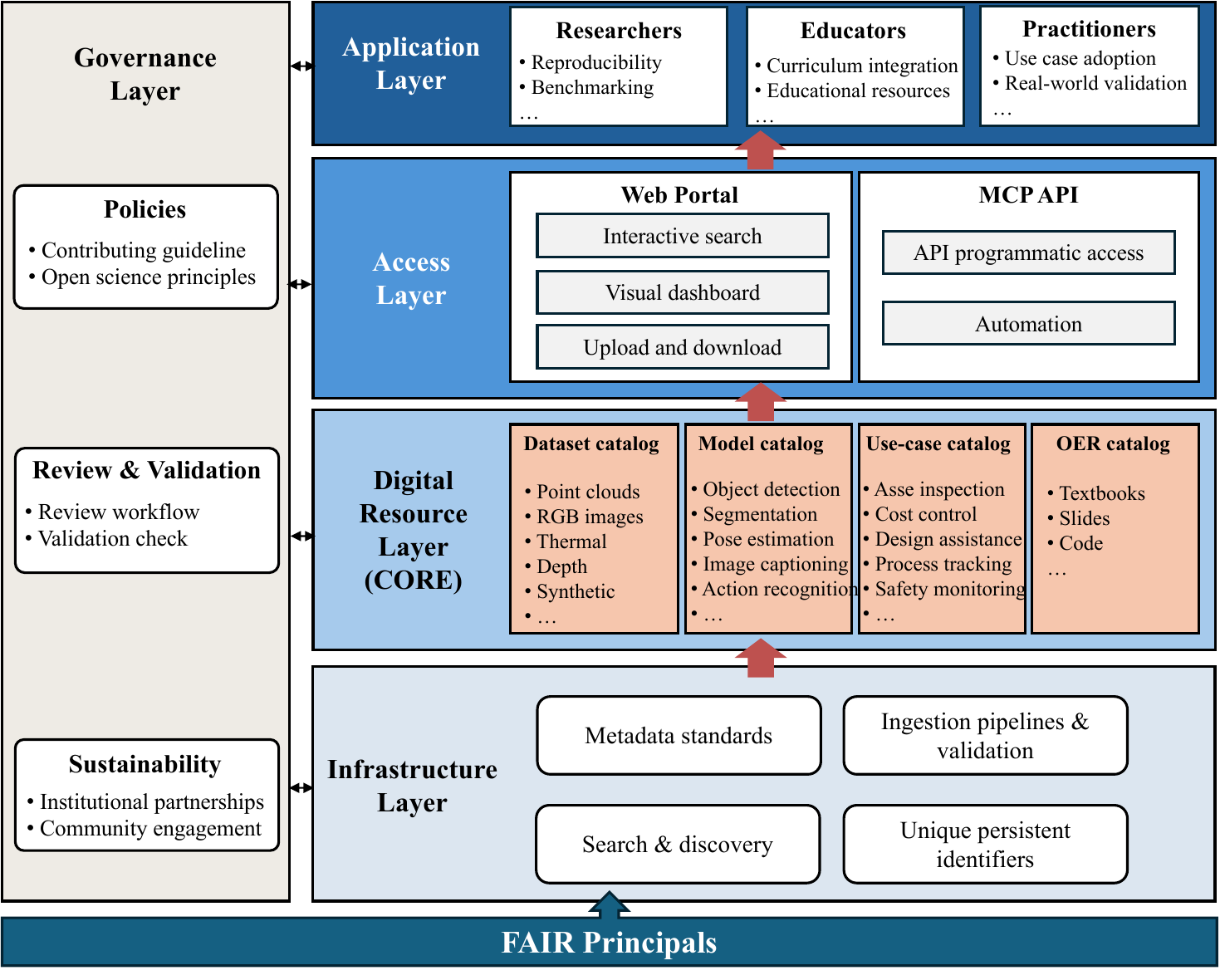}
    \caption{Conceptual framework of the \textsc{OpenConstruction} platform}
    \label{fig:architecture}
\end{figure}

This study presents the conceptual framework of the \textsc{OpenConstruction} platform, focusing on the design principles that guide the organization and governance of open AEC digital resources. As illustrated in Fig.~\ref{fig:architecture}, the platform integrates five interrelated layers that operationalize the FAIR principles~\cite{wilkinson2016fair} to allow collaborative knowledge sharing.

\subsection{Vision and Guiding Principles}
The vision of the \textsc{OpenConstruction} platform is to transform fragmented digital resources into shared community assets and reusable AEC knowledge. Guided by FAIR principles \cite{wilkinson2016fair}, the ecosystem emphasizes persistent identifiers, standardized metadata, and validation workflows to ensure that datasets, models, use cases, and educational resources can be discovered, interpreted, and reused.

These principles are implemented through coordinated technical and organizational mechanisms. Standardized metadata schemas and descriptors provide semantic consistency, while ingestion and validation pipelines verify schema completeness, format integrity, and license compliance. The access architecture integrates interactive visualization, semantic search, and API integration to support human exploration and machine-driven discovery.

\subsection{Architecture of \textsc{OpenConstruction} Platform}
The \textsc{OpenConstruction} architecture comprises five layers that collectively define how digital resources are stored, organized, accessed, and governed within the platform.

\subsubsection{Infrastructure Layer: Technical Foundation}
The infrastructure layer establishes the foundational technical environment for metadata management and interoperability. Instead of storing data or models directly, the platform maintains federated metadata records that reference externally hosted resources curated by their original authors. Each record is assigned a persistent identifier and structured according to a unified metadata schema that captures key descriptors such as modality, scope, licensing, and provenance. Standardized metadata extraction and compliance checks ensure standardization across diverse sources. The infrastructure also supports semantic search, link validation, and provenance visualization, enabling transparent verification of digital resources. All metadata ingestion and update operations use authenticated contributor accounts and secure transmission protocols, preventing unauthorized modification.

\subsubsection{Service Layer: Knowledge Organization}
Based on this foundation, the service layer organizes the indexed information into four interlinked catalogs that structure the open science platform. The \textit{dataset catalog} aggregates metadata from multimodal datasets, such as images and point clouds, extracted from open-access repositories and archives. The \textit{model catalog} indexes computational models for AEC tasks, including safety monitoring and design optimization. The \textit{use-case catalog} documents validated applications of AI and automation in design, construction, and maintenance, connecting practical deployments and observed outcomes. The \textit{OER hub} gathers open instructional materials and promotes community engagement.

\subsubsection{Access Layer: Interoperability Interfaces}
The access layer mediates the interaction between users, computational agents, and the federated catalogs through graphical and programmatic interfaces. The Web portal provides an interactive environment for metadata-driven search, visualization, and contribution, allowing users to explore resources through unified descriptors and provenance records. In parallel, the Model Context Protocol (MCP)~\cite{hou2025model} offers a machine-actionable interface that supports automated retrieval, validation, and integration of metadata via standardized APIs. This architecture extends FAIR compliance from human-readable access to machine interpretability, allowing AI systems and analytical pipelines to query and connect heterogeneous external resources in a reproducible manner. 
 
\subsubsection{Application Layer: Research, Education, and Practice}
The application layer represents the primary domains where the indexed resources are applied and reused. For researchers, the platform facilitates cross-dataset benchmarking, reproducibility assessments, and transparent sharing of computational workflows. For educators, this platform provides curated open materials for teaching data-centric methods, AI literacy, and digital AEC practices. For practitioners, this platform serves as a reference environment for validated use cases and open-source tools that support technology adoption and knowledge transfer. By connecting indexed resources to practical contexts, this layer enables users to interpret federated metadata into actionable knowledge that enhances research transparency, pedagogical innovation, and industry implementation.

\subsubsection{Governance Layer: Community and Sustainability}
The governance layer defines the mechanisms that ensure transparency, accountability, and long-term sustainability. All submissions undergo manual validation for metadata completeness, accessibility, and ethical compliance. Contributors are publicly acknowledged on the website through visible profiles and institutional affiliations, promoting transparency and community participation. Community involvement is promoted through regular updates, subscription-based notifications, and workshops that encourage knowledge exchange and collaboration between research and industry stakeholders.

\section{Platform Implementation and Workflows}
\begin{figure*}
    \centering
    \includegraphics[width=0.9\linewidth]{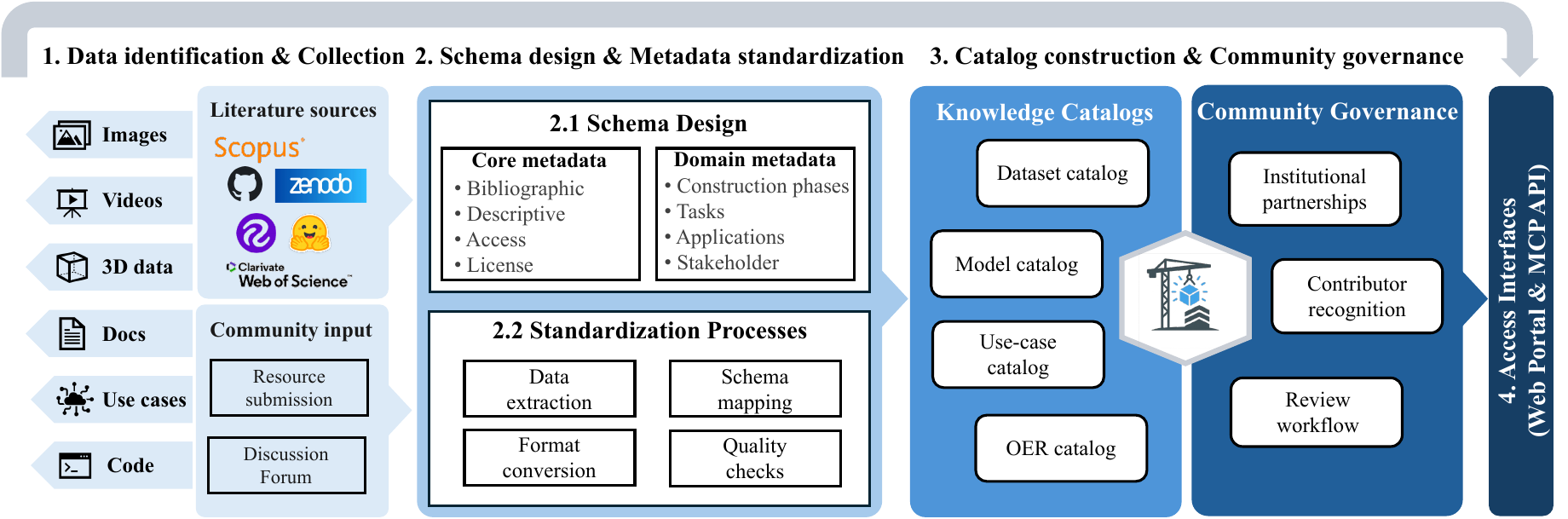}
    \caption{Implementation workflow of the \textsc{OpenConstruction} open science platform}
    \label{fig:workflow}
\end{figure*}

Building on the conceptual framework presented in Section \ref{sec:conceptual}, this section describes the concrete implementation of the \textsc{OpenConstruction} platform, detailing metadata schemas, ingestion pipelines, validation procedures, and community workflows (see Fig.~\ref{fig:workflow}).

\subsection{Data Identification and Collection}
Resource identification compile open \emph{datasets}, \emph{models}, \emph{use cases}, and \emph{OERs} in AEC. Three keyword categories were combined using Boolean \texttt{AND}: (1) \textit{domain context} (e.g., “AEC,” “construction management”), (2) \textit{method terms} (e.g., “artificial intelligence,” “machine learning,” “large language models”), and (3) \textit{access and application descriptors} (e.g., “open access,” “dataset,” “model,” “case study”).

Searches covered academic databases (Scopus, Web of Science), open-data repositories (Zenodo, Roboflow, Google Dataset), and code repositories (GitHub, Hugging Face). \emph{Use cases} were identified through targeted Google searches (e.g., “AI use cases in AEC”), with the top 200 ranked results per query manually screened. \emph{OERs} were collected by searching established open-education platforms, such as Pressbooks, for openly accessible materials relevant to AEC education.

All retrieved records were screened and merged in two stages (title/abstract and full review). Entries were included if they (1) addressed AEC-related tasks, (2) provided open resources, and (3) contained referenced documents. Proprietary, restricted, or non-AEC resources were excluded. Table~\ref{tab:search} summarizes the data sources and inclusion criteria.
\begin{table*}[ht]
\centering
\caption{Catalog types, keyword categories, and inclusion/exclusion criteria used for curating \textsc{OpenConstruction} platform.}
\label{tab:search}
\scriptsize
\resizebox{\linewidth}{!}{%
\begin{tabular}{p{1.5cm}p{2.5cm}p{4cm}p{4.1cm}p{2.8cm}}
\toprule
\textbf{Catalog Type} & \textbf{Sources} & \textbf{Keyword Categories / Search Method} & \textbf{Inclusion Criteria} & \textbf{Exclusion Criteria} \\
\midrule

Dataset Catalog &
Scopus; Web of Science; Zenodo; Roboflow; Google Dataset Search; GitHub; Hugging Face &
(1) “AEC,” “construction,” “built environment”; 
(2) “artificial intelligence,” “machine learning,” “deep learning,” “computer vision”; 
(3) “dataset,” “benchmark,” “open access” &
Publicly available datasets relevant to AEC tasks with open or academic licenses &
Non-AEC domains; proprietary or restricted datasets; inaccessible or invalid links \\

Model Catalog &
Scopus; Web of Science; GitHub; Hugging Face &
(1) “AEC,” “construction,” “built environment”; 
(2) “artificial intelligence,” “machine learning,” “deep learning”; 
(3) “model,” “algorithm,” “code repository” &
Open-source models or codebases applicable to AEC tasks, with accessible documentation &
Non-AEC models; proprietary or unavailable code; repositories lacking usable documentation \\

Use Case Catalog &
Google Search (manual curation) &
Queries such as “AI use cases in AEC,” “AI in construction practice”; top 200 results per query screened &
Documented real-world or field-tested AI/ML applications in AEC from credible sources &
Non-AEC applications; purely conceptual descriptions; unverifiable claims \\

OER Catalog &
Open-education platforms (e.g., Pressbooks) &
Searches for AEC-relevant open textbooks, tutorials, and training materials &
Openly accessible educational resources supporting AEC teaching &
Paywalled materials; proprietary training content; irrelevant subject areas \\

\bottomrule
\end{tabular}}
\end{table*}

\subsection{Schema Design and Metadata Standardization}
The metadata schema builds on our previous studies on organizing construction datasets and computational workflows \cite{wang2023characterizing,xiong2025openconstruction}. The schema includes (1) core descriptors, including identifier, title, contributors, license, access link, and (2) domain-specific descriptors, such as project phases, tasks, applications, stakeholders, and technologies.

Metadata standardization in the platform follows a lightweight process focused on consistency. Metadata fields are first extracted from source repositories using their accompanying documentation or repository files. These fields are then aligned with the platform schema through manual inspection to ensure that heterogeneous records use consistent descriptors and terminology. Basic format checks are performed through simple scripts, while final validation is conducted manually to confirm completeness, licensing information, and appropriate use of controlled vocabularies for modalities, tasks, and applications. This process enables consistent representation of diverse resources and supports cross-catalog linkage and transparent provenance tracking.

\subsection{Catalog Construction and Ingestion Pipelines}
Catalog construction follows an integrated ingestion pipeline that combines systematic harvesting from external repositories with community-driven submissions. Harvesting routines periodically retrieve open-access datasets, models, and OERs from sources such as Scopus, Zenodo, and GitHub. The extracted metadata are normalized using the standardized schema and unique identifiers. Community contributors can submit additional resources through an online form aligned with the same schema, enabling consistent metadata capture.

Each submission or harvested record undergoes manual validation to verify field completeness and link accessibility. Potential duplicates are detected through title, author, and repository matching. Entries that meet the completeness and relevance requirements are reviewed for quality assurance and ethical compliance. Approved resources are indexed and published in the live catalog through automated synchronization.

\subsection{Community Services and Access}
Community services  enable users to contribute, explore, and connect resources through structured submission workflows, discussion forums, and contributor dashboards. Standardized templates ensure metadata consistency, while interlinked catalogs and update notifications support collaboration and feedback. The platform is accessible through two interoperable interfaces: a user-oriented web portal and machine-actionable MCP APIs \cite{hou2025model}. The Web portal provides interactive search, filtering, and visualization, while the MCP server enables programmatic retrieval and integration of catalog entries into research and educational workflows.

\section{Results}
This section reports on the current scope of catalogs and illustrates their utility through case studies.

\subsection{Scope of the \textsc{OpenConstruction} Ecosystem}
\begin{figure}[htbp]
    \centering
    \includegraphics[width=.7\linewidth]{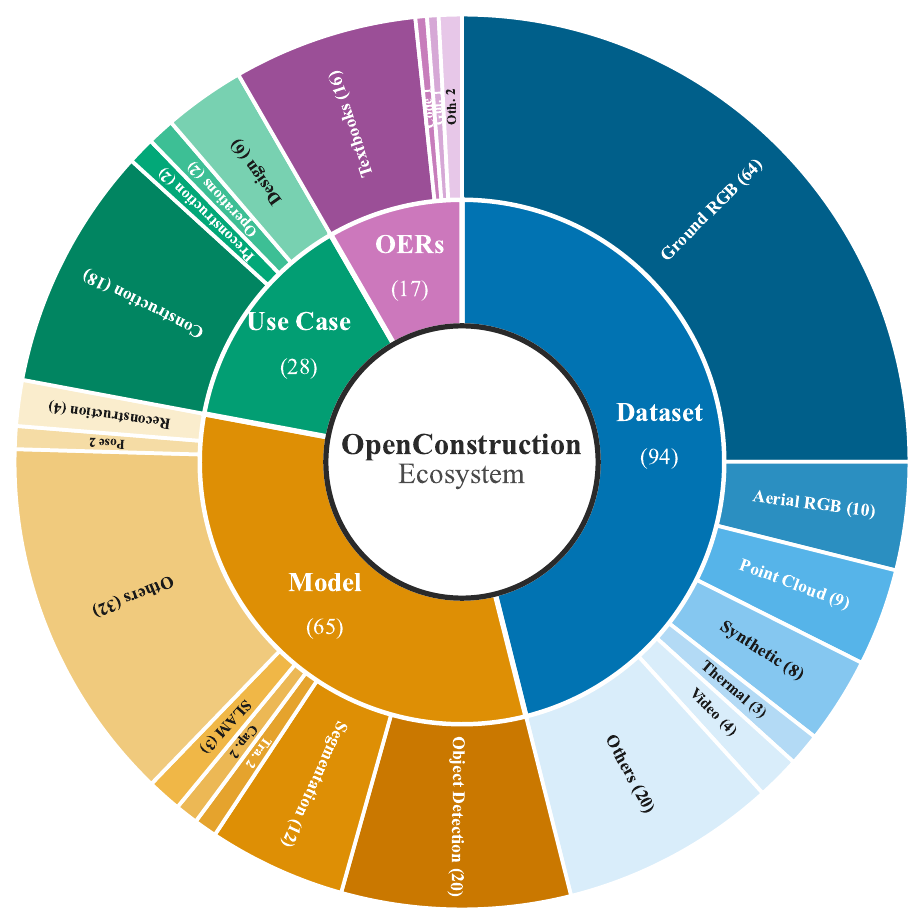}
    \caption{Distributions of the \textsc{OpenConstruction} ecosystem. The inner ring shows catalog types, and the outer ring illustrates category proportions. Note that individual datasets and models may belong to multiple categories.}
    \label{fig:data}
\end{figure}

As of December 2025, the \textsc{OpenConstruction} platform hosts 204 entries in four catalogs (Fig.~\ref{fig:data}). The \textit{Dataset Catalog} (94 entries) covers multiple modalities: ground-level RGB imagery dominates (64 datasets), followed by aerial RGB (10), point clouds (9), synthetic data (8), and modalities such as thermal and video datasets. The \textit{Model Catalog} (65 entries) is led by object detection models (20). Additional entries include segmentation (12), tracking (2), pose estimation (2), SLAM (3), image captioning (2), and 3D reconstruction models (4). The \textit{Use Case Catalog} (28 entries) documents real-world implementations, with 18 cases from construction, 2 from preconstruction, 2 from operations and maintenance, and 6 from design phases. The \textit{OER Catalog} (17 entries) provides openly accessible textbooks (16) and slides (1) on AEC, computing, and data-intensive topics to support education and training.

\subsection{Case Studies}
This section presents two case studies outlining the use of the proposed platform in research-related exploration and in instructional activities.

\subsubsection{Case Study 1: Model Discovery and Cross-Study Comparison}
\begin{figure}[ht]
\centering
\subfigure[Web portal for model discovery and metadata browsing \label{fig:web}]
{\includegraphics[width=0.8\linewidth]{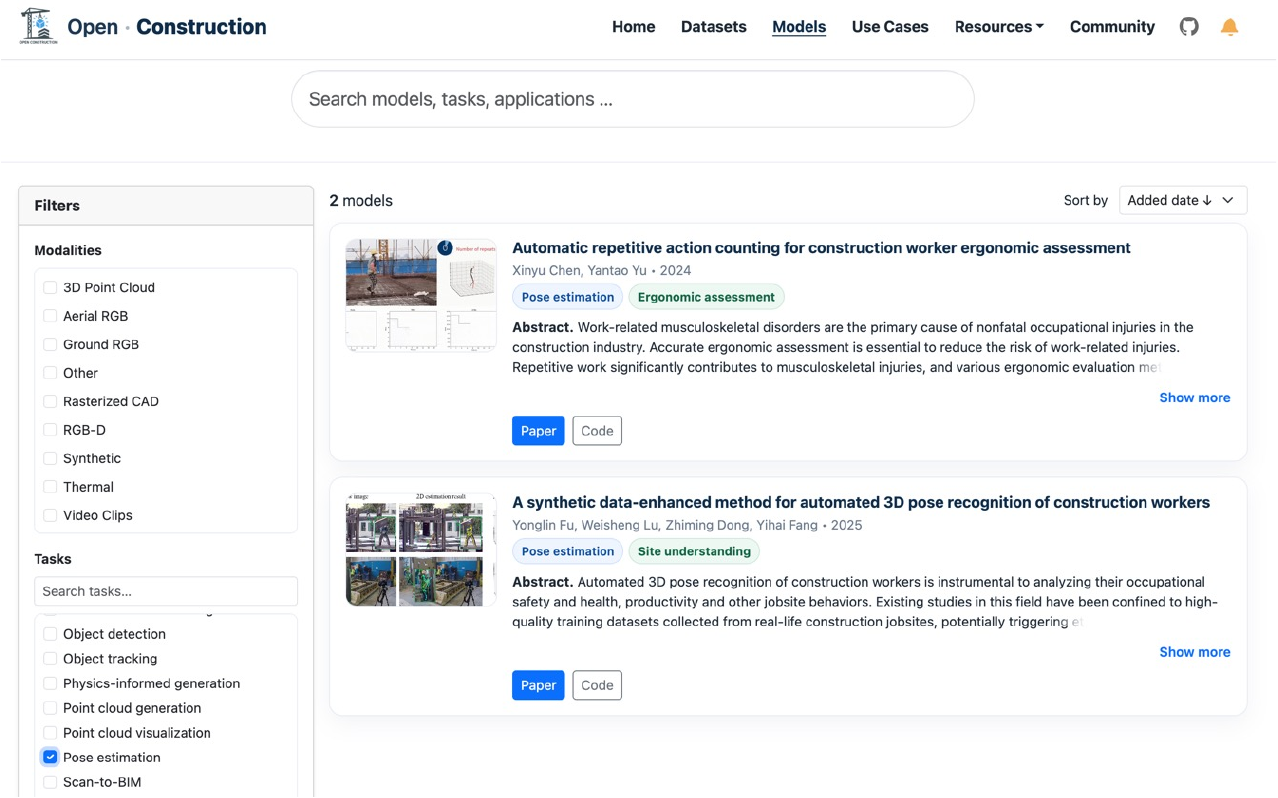}}
\subfigure[MCP server for querying, aggregation, and visualization \label{fig:mcp}]
{\includegraphics[width=0.8\linewidth]{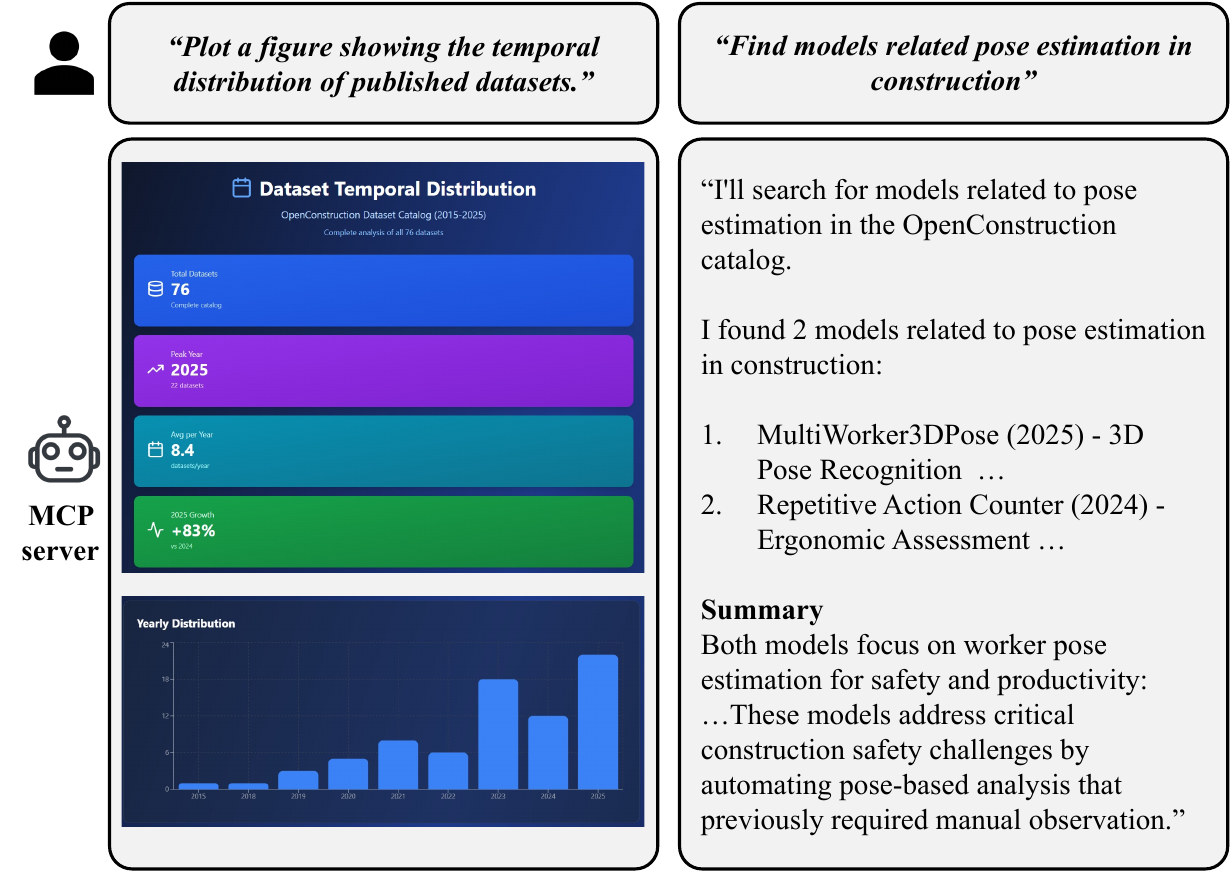}}
\caption{Integrated workflow for model discovery and benchmarking}
\label{fig:access}
\end{figure}

The platform provides a unified environment for identifying models and their associated datasets across AEC tasks. Through the web portal (Fig.~\ref{fig:web}), users can filter models by task, modality, and application. For instance, a query for “pose estimation” retrieves models such as \emph{MultiWorker3DPose} (2025) and \emph{Repetitive Action Counter} (2024), each accompanied by standardized descriptors including task definitions, modality, license, and links to datasets and documentation. The MCP endpoint (Fig.~\ref{fig:mcp}) provides the same metadata in machine-readable form, enabling scripted retrieval and aggregation of model attributes for analysis or integration into research pipelines. This case study illustrates how the platform consolidates model documentation and supports consistent comparison across independently developed resources.

\subsubsection{Case Study 2: Instructional Use in AEC Education}
The platform resources can be incorporated into instructional settings as a reference environment for examining curated AEC datasets, models, and metadata. Learners can query catalog entries to compare sensing modalities, annotation structures, and task definitions, and review model records to understand declared inputs, outputs, and application domains. For example, a learner may retrieve multiple construction-safety image datasets to examine differences in annotation formats and then inspect the corresponding detection models to see how these resources are linked through standardized descriptors.

\section{Discussion}
\label{sec:discuss}
This section examines implications of the proposed cyberinfrastructure for research, education, and practice in AEC, and identifies challenges that shape future development.

\subsection{Implications for Research, Education, and Practice}
The proposed platform can support the collaborative framework and reduce fragmentation in AEC digital resources. For research, standardized descriptors and systematic curation improve discoverability, transparency, and cross-study comparison, directly supporting the cumulative development of data-driven methods. In education, curated catalogs and OERs provide students with access to reproducible datasets and models, strengthening computational literacy, and linking research methods to professional practice. For practice, the platform enables practitioners to evaluate emerging solutions using validated datasets and models. 

\subsection{Ethical, Legal, and Privacy Considerations}
Open sharing of AEC resources raises concerns related to privacy, proprietary information, and intellectual property. Datasets may contain identifiable individuals or sensitive operational details. \textsc{OpenConstruction} platform mitigates these risks through required license declarations, provenance documentation, and a curator review workflow. However, community-wide agreements on anonymization practices, responsible AI use, and data-sharing policies remain necessary to ensure ethical and legally compliant knowledge dissemination.

\subsection{Limitations and Future Works}
The current platform demonstrates the feasibility of an AEC-specific cyberinfrastructure, but several limitations remain. Catalogs are still dominated by visual and sensing resources, while domains such as scheduling and costs are underrepresented. The underlying design of the platform, built around extensible catalogs and schema-based metadata, can easily incorporate these missing domains as new resource types and descriptors are introduced. Sustaining community contribution is another challenge, as researchers and practitioners may be reluctant to share resources without clear recognition \cite{wang2023characterizing}. Implementing contribution metrics may improve engagement. Long-term sustainability will depend on institutional support and community governance, with professional societies playing a key coordinating role.

\section{Conclusion}
\label{sec:conclude}
This study introduced \textsc{OpenConstruction} platform, a domain-specific cyberinfrastructure that consolidates datasets, models, use cases, and tools to advance knowledge sharing in the AEC sector. By embedding workflows for metadata standardization, ingestion, and validation, the platform operationalizes the FAIR principles and addresses persistent challenges of data fragmentation, inconsistent documentation, and limited interoperability. The case studies demonstrate that this open platform can transform fragmented digital assets into a structured, cumulative, and verifiable knowledge base, enabling cross-study benchmarking for model evaluation and computational training in AEC education.

However, scaling such systems requires sustainable governance, ethical and legal safeguards for data sharing, and contributor recognition mechanisms that incentivize participation. As long-term sustainability and quality assurance are foundational to the advancement of the AEC community, our objective is to foster the participatory practices needed to support continuity as the platform evolves. The platform establishes an open-science infrastructure that supports transparent dissemination of community-curated resources and cumulative innovation across research and practice.

\vspace{5mm}
\noindent\textbf{{Acknowledgments}}
We sincerely thank all creators and contributors who have shared datasets, models, and other resources with the community. This research was supported by the Farris Family Innovation Award and the Cajun Industries Professorship. The findings, interpretations, and conclusions expressed in this study do not necessarily reflect the views of Farris Family or Cajun Industries.

\appendix

 \bibliographystyle{elsarticle-num} 
 \bibliography{cas-refs}





% \end{thebibliography}
\end{document}